# Exact counting of Euler Tours for generalized series-parallel graphs

Prasad Chebolu[*]   Mary Cryan[†]   Russell Martin[*]

**Abstract**

We give a simple polynomial-time algorithm to exactly count the number of Euler Tours (ETs) of any Eulerian generalized series-parallel graph, and show how to adapt this algorithm to exactly sample a random ET of the given generalized series-parallel graph. Note that the class of generalized series-parallel graphs includes all outerplanar graphs. We can perform the counting in time $\mathcal{O}(m\Delta^3)$, where $\Delta$ is the maximum degree of the graph with $m$ edges. We use $\mathcal{O}(m\Delta^2 \log \Delta)$ bits to store intermediate values during our computations. To date, these are the first known polynomial-time algorithms to count or sample ETs of any class of graphs; there are no other known polynomial-time algorithms to even approximately count or sample ETs of any other class of graphs. The problem of counting ETs is known to be $\sharp P$-complete for general graphs (Brightwell and Winkler, 2005 [3]) and also for planar graphs (Creed, 2009 [4]).

[*]Department of Computer Science, University of Liverpool, Ashton Bldg, Liverpool L69 3BX, UK. Supported by EPSRC grant EP/F020651/1.

[†]Lab for Foundations of Computer Science, School of Informatics, University of Edinburgh, Edinburgh EH8 9AB Scotland, UK. Supported by EPSRC grant EP/D043905/1.

# 1 Introduction

Let $G = (V, E)$ denote an undirected, connected multigraph where the degree, $d(v)$, of each vertex $v \in V$ is even. Any standard introductory graph theory text has a result stating that every graph with even degree has an *Euler tour* (or *Euler circuit*), i.e., a circuit that traverses every edge of $G$ exactly once. This result, which also holds for Eulerian multigraphs, implies a very simple linear-time algorithm for testing whether a given multigraph admits some Euler tour. In this paper we consider the counting and sampling of Euler tours of generalised series-parallel graphs, a special class of multigraphs. Throughout the paper we will use the term 'graph' to include multigraphs (ie, to allow the possibility of loops and parallel edges).

In 2005, Brightwell and Winkler showed that the problem of counting Euler tours is $\sharp P$-complete[3]. This is in sharp contrast to the case of *directed* Eulerian graphs (i.e. connected digraphs for which the indegree equals the outdegree at each vertex), where the number of Euler tours can be counted exactly in polynomial-time using the Matrix-Tree Theorem [2] and the so-called "BEST" Theorem (after de **B**ruijn, van Aadenne-**E**hrenfest, **S**mith, and **T**utte, although apparently the first two deserve credit as the original discoverers [1]). More recently, Creed [4] showed that counting the number of Euler tours in undirected graphs remains $\sharp P$-complete if $G$ is restricted to be a planar graph.

These $\sharp P$-completeness results naturally lead one to question for which classes of graphs can (exact or approximate) counting of the number of Euler tours be done efficiently. We consider the case of *generalized series-parallel (GSP)* graphs, a subclass of planar graphs with distinguished source $s$ and sink $t$ vertices, that may be constructed in an inductive manner using a small number of operations. The key operations combine two generalized series-parallel graphs $G_1, G_2$ to form a larger graph, and these operations are known as series composition $G_1 o_s G_2$, parallel composition $G_1 o_p G_2$, and the dangling composition $G_1 o_d G_2$. These operations are all defined in Section 2.1. The problem of checking whether a given graph $G$ is a generalized series-parallel graph can be done in polynomial time [6], and if $G$ is a GSP graph, a hierarchical binary tree decomposition of $G$ (see Section 2.1) can be found in polynomial time [5, 12, 13, 14, 10, 6]. In a hierarchical tree decomposition $T$ of a generalized series-parallel graph $G$, each internal node $u$ is associated with an operation $o_u$ (one of $o_s$, $o_p$ and $o_d$) and each leaf node $u$ is associated with an edge (of $G$). For every internal node $u$ of the tree decomposition, with child nodes $v$ and $w$, the subtree $T_u$ represents a connected subgraph $G_u$ of $G$, where $o_u$ is the top-level operator combining $G_v$ and $G_w$.

In this paper we give a polynomial-time "dynamic-programming"-like algorithm for exactly counting Euler tours for GSP graphs. Our main result is as follows:

**Theorem 1** *Let $G$ be an Eulerian generalized series-parallel graph having $m$ edges and maximum degree $\Delta$. We assume that we know $T$, a binary tree decomposition for $G$.*

*Counting the number of Euler tours of $G$ can be performed using $\mathcal{O}(m\Delta^3)$ arithmetic operations, and using $\mathcal{O}(m\Delta^2 \log \Delta)$ bits for storing intermediate values in the computations.*

We also show how we can use our results to sample an Euler tour of a GSP graph exactly uniformly at random. To the best of our knowledge, these are the first polynomial-time algorithms for counting (and sampling uniformly at random from) the set of Euler tours for any class of undirected graphs. At the time of writing, we do not even know of any approximate-counting results for any significant class of undirected graphs. One of the few relevant results of which we are aware is an approximate asymptotic formula for the number of Euler tours of the complete graph on any odd number of vertices [8].

For general graphs, there exist natural Markov chains for sampling Euler tours such as the "Kotzig chain", which uses circuit reversals. However, although there has been some research on the Kotzig Markov chain [11], no correct proof of rapid mixing has yet been found for any class of Eulerian graphs.

Our paper is structured as follows: in Section 2 we give some key definitions, including the definition of an $(s, t)$-decomposition of an Eulerian graph with distinguished vertices $s, t$. Most of the work in this paper deals with the relationship between these $(s,t)$-decompositions and Euler tours, and the building-up of counts of these $(s,t)$-decompositions. In Section 3 we will show that the number of Euler tours of a Eulerian graph $G$ can be expressed as a simple weighted sum over the count of $(s,t)$-decompositions with $k$ non-loop paths for a linear number of $k$ values. Therefore the main component of both our counting and sampling algorithms will be to build a table containing the counts of $(s,t)$-decompositions, for all component graphs of $G$, and all relevant $k$. In Section 4 we show how the counts of $(s,t)$-decompositions of two GSP graphs $G_1, G_2$ (for varying values of $k$) can be used to build the counts for the parallel-composition $G_1 o_p G_2$, the series-composition $G_1 o_s G_2$ and the dangling-composition $G_1 o_d G_2$. Finally in Section 5 we sketch the simple polynomial-time algorithms that allow us to exactly count and sample Euler tours of GSP graphs.

## 2 Definitions

Throughout this paper we assume that $G = (V, E)$ is a multigraph, which may contain parallel edges and loop edges. We will assume that the edges of the graph have some arbitrary but fixed ordering $e_1, \ldots, e_m$ (where $m$ is the number of edges of the graph). We will use $Adj_G(v)$ to denote the set of edges adjacent to the vertex $v$. We also use the notation $\mathbb{N}_0 = \{0, 1, 2, \ldots\}$.

### 2.1 Generalized series-parallel graphs

First we define the class of graphs that we study in this paper.

**Definition 2.1** *A* generalized series-parallel graph (GSP graph) *is any graph $G = (V, E, s, t)$ with two distinguished nodes $s, t \in V$, which can be built inductively from the following operations:*

- *B: The graph consisting of two vertices connected by a single edge is a GSP graph (where $s$ and $t$ are the endpoints of the single edge).*

- $o_s$: *Given two GSP graphs $G_1 = (V_1, E_1, s, t)$, $G_2 = (V_2, E_2, s', t')$, the* series composition *of $G_1, G_2$ is defined as $G = G_1 o_s G_2 =_{def} (V, E, s, t')$, where $V =_{def} (V_1 \cup V_2) \setminus \{s'\}$ and $E =_{def} (E_1 \cup E_2 \cup \{(t, v) : \forall e = (s', v) \in Adj_{G_2}(s')\}) \setminus Adj_{G_2}(s')$.*

- $o_p$: *Given two GSP graphs $G_1 = (V_1, E_1, s, t)$, $G_2 = (V_2, E_2, s', t')$, the* parallel composition *of $G_1, G_2$ is defined as $G = G_1 o_p G_2 = (V, E, s, t)$, where $V =_{def} (V_1 \cup V_2) \setminus \{s', t'\}$, and $E =_{def} (E_1 \cup E_2 \cup \{e' = (s, v) : \forall e = (s', v) \in Adj_{G_2}(s')\} \cup \{e' = (t, v) : \forall e = (t', v) \in Adj_{G_2}(t')\}) \setminus (Adj_{G_2}(s') \cup Adj_{G_2}(t'))$.*

- $o_d$: *Given two GSP graphs $G_1 = (V_1, E_1, s, t)$, $G_2 = (V_2, E_2, s', t')$, the* dangling composition *of $G_1, G_2$ is defined as $G = G_1 o_d G_2 = (V, E, s, t)$, where $V =_{def} (V_1 \cup V_2) \setminus \{s'\}$, and $E =_{def} (E_1 \cup E_2 \cup \{e' = (s, v) : \forall e = (s', v) \in Adj_{G_2}(s')\}) \setminus Adj_{G_2}(s')$.*

Intuitively, the series-composition of $G_1$ and $G_2$ is formed by identifying the sink $t$ of $G_1$ with the source $s'$ of $G_2$, with the new source and sink of $G_1 o_s G_2$ being, respectively, the source of $G_1$ and the sink of $G_2$. The parallel-composition of $G_1$ and $G_2$ is formed by identifying the source $s$ of $G_1$ with the source $s'$ of $G_2$, and the sink $t$ of $G_1$ with the sink $t'$ of $G_2$. The dangling-composition of $G_1$ and $G_2$ is formed by identifying the source $s$ of $G_1$ with the source $s'$ of $G_2$ (and keeping the source and sink of $G_1$). Note that the order of $G_1, G_2$ is important for the $o_s$ and $o_d$ operations. Moreover, $G_1 o_s G_2$ and $G_2 o_s G_1$ need not be isomorphic to one another.

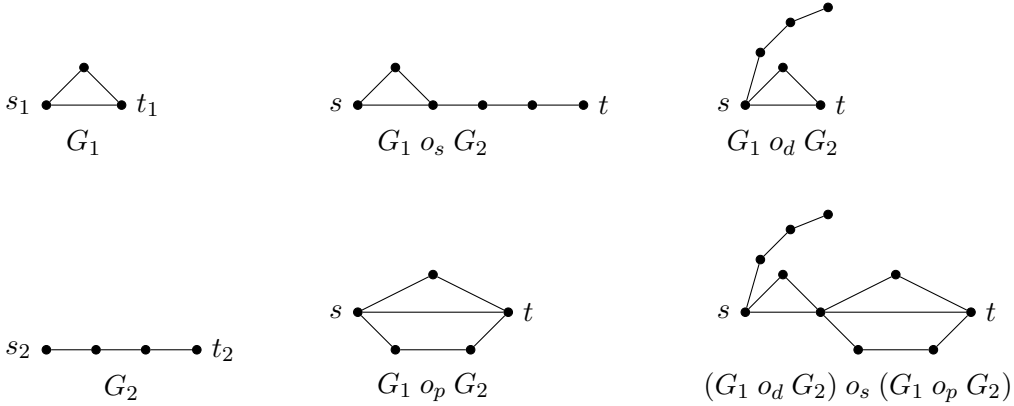

Figure 1: Examples of the operations for constructing GSP graphs.

The class of series-parallel graphs consists of those that may be obtained using the three operations $B, o_s$, and $o_p$. Adding the dangling-composition operation brings us to the class of generalized series-parallel graphs. Out of interest we note that outerplanar graphs are known to be generalized series-parallel graphs [7]. Figure 1 shows some examples of the operations used to construct GSP graphs.

We note here that GSP graphs have a compact representation using a binary tree. Each leaf of the binary tree is a (labelled) edge of the graph, and each internal vertex of the tree represents a series, parallel, or dangling operation that applies to its two children, and each internal vertex is labelled appropriately. In the case of a series (resp. dangling) operation, we can define the tree so that the left child corresponds to $G_1$ and the right one $G_2$ in the definition of the series (resp. dangling) operation. See Figure 2 for an example. We haven't explicitly indicated the source and sink node for each of the subgraphs, but they are obvious for this small example.

## 2.2 Euler tours and "legal" graphs

**Definition 2.2** *A connected graph is said to be* Eulerian *if every vertex has even degree. A graph is said to be* near-Eulerian *(with discrepancies at $v, v'$) if all but two vertices $v, v'$ have even degree. We will say that a graph $G = (V, E, s, t)$ with distinguished vertices $s, t$ is* legal *if the graph is either Eulerian or is near-Eulerian with discrepancies at $s$ and $t$.*

The particular class of legal graphs we will focus on in Sections 4 and 5 are GSP graphs. The four operations B, $o_s$, $o_p$ and $o_d$ for constructing GSP graphs only change the degree of the distinguished vertices $s, t$. Therefore, in building a GSP graph which is Eulerian, we need only consider graphs which are either Eulerian, or are near-Eulerian with discrepancies at $s$ and $t$. Note that any graph constructed by the base-case operation B is by default a legal graph. A series-composed graph $G = G_1 o_s G_2$ is a legal graph if and only the following conditions hold:

s1. $G_1$ and $G_2$ are both legal graphs, and

s2. either $d_{G_1}(t), d_{G_2}(s')$ are both even, or $d_{G_1}(t), d_{G_2}(s')$ are both odd.

A parallel-composed graph $G = G_1 o_p G_2$ is a legal graph if and only if:

p1. $G_1$ and $G_2$ are both legal graphs.

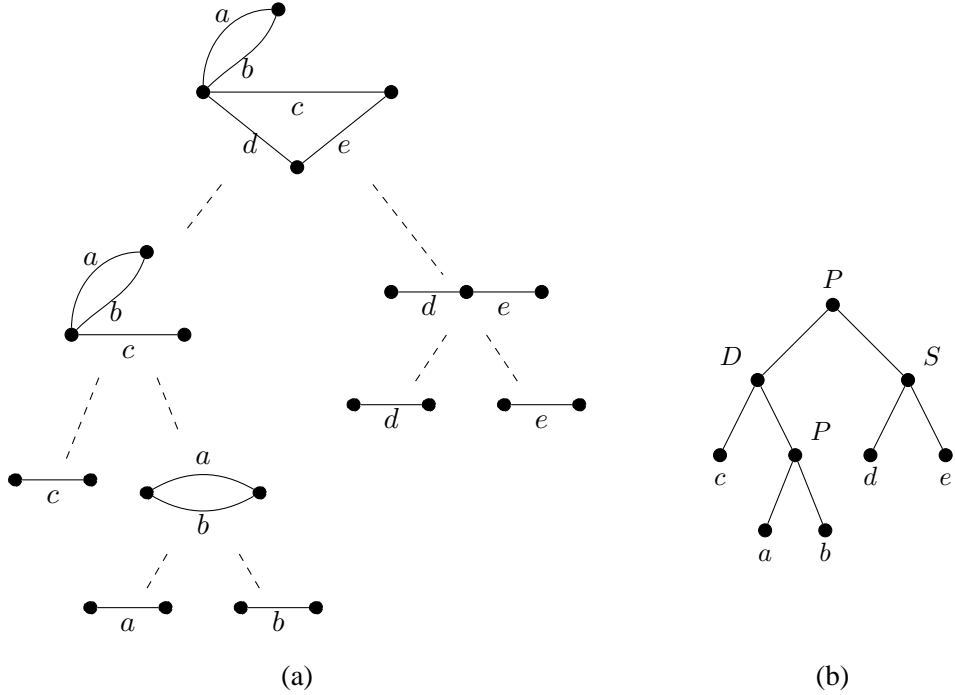

Figure 2: A GSP graph construction in (a), and its corresponding binary tree decomposition in (b).

A dangling-composed graph $G = G_1 \circ_d G_2$ is a legal graph if and only if:

d1. $G_1$ is a legal graph and $G_2$ is an Eulerian legal graph.

Note that for any legal graph, the parity of the source is the same as the parity of the sink.

**Observation 2** *Let $G = (V, E, s, t)$ be any legal GSP graph, and $T$ any (rooted) binary tree decomposition of $G$ according to $o_p$, $o_s$ and $o_d$. Then for every node $u$ of $T$, the subgraph $G_u$ corresponding to the subtree $T_u$ is a legal GSP graph.*

**Definition 2.3** *For any Eulerian graph $G$, an* Euler tour *$T$ is any path $T$ in $G$ which traverses every edge exactly once. We consider Euler tours to be identified under the operations of rotation and reversal. For any graph $G$, we let $ET(G)$ denote the set of Euler Tours of $G$.*

**Definition 2.4** *Let $G = (V, E, s, t)$ be a legal graph, and let $p$ be a (not necessarily simple) path in $G$. We say that $p$ is a $(s,t)$-*simple *path if one endpoint of $p$ is $s$ and the other is $t$, but none of the intermediate points of $p$ lie in $\{s, t\}$. An $(s,s)$-*loop *is a (not necessarily simple) circuit that starts and ends at $s$, such that neither $s$ nor $t$ are intermediate nodes of the circuit. A $(t,t)$-*loop *is defined similarly.*

Note that an $(s,t)$-simple path, an $(s,s)$-loop or a $(t,t)$-loop may visit vertices of $V \setminus \{s, t\}$ more than once.

**Definition 2.5** *Let $G = (V, E, s, t)$ be a legal graph. An $(s,t)$-*decomposition *of $G$ is any collection $C = \{p_1, \ldots, p_\ell\}$ of edge-disjoint $(s,t)$-simple paths, $(s,s)$-loops, and $(t,t)$-loops in $G$ such that $\cup_{i=1}^{\ell} \{e : e \in p_i\} = E$, and such that for every $i \in [\ell]$, the initial edge of $p_i$ has an edge label with a lower index than that of the final edge of $p_i$.*

Informally, an $(s,t)$-decomposition is a partition of the edges of $G$ into $(s,t)$-simple paths, and loops that contain either $s$ or $t$, but not both. Any Euler tour of $G$ gives rise to a unique $(s,t)$-decomposition in a natural way. Conversely, any fixed $(s,t)$-decomposition will give rise to a number of Euler tours in $G$ (see Lemma 6 below). We have the following observation.

**Observation 3** *Let $G = (V, E, s, t)$ be a legal graph. If $C$ is an $(s,t)$-decomposition of $G$, then $\ell = |C| = \frac{d(s)+d(t)}{2}$.*

## 3 $(s,t)$-decompositions and Euler Tours

In this section we demonstrate the relationship between the set of $(s,t)$-decompositions of an Eulerian graph (with distinguished vertices $s, t$) and the set of Euler tours of that graph.

**Definition 3.1** *Let $G = (V, E, s, t)$ be an Eulerian graph with distinguished vertices $s$ and $t$, and let $T \in ET(G)$. We say that the $(s,t)$-decomposition $C$ of $G$ is consistent with the Euler tour $T$ if for every $p \in C$, either $p$ or $rev(p)$ is a contiguous segment of $T$ (where $rev(p)$ is the reverse of the path $p$).*

**Observation 4** *Let $G = (V, E, s, t)$ be an Eulerian graph with distinguished vertices $s$ and $t$ and suppose $T \in ET(G)$. Then there is exactly one $(s,t)$-decomposition of $G$ which is consistent with $T$.*

We make the following definition for all legal graphs.

**Definition 3.2** *Let $G = (V, E, s, t)$ be an legal graph. Let*

$$\kappa(G) = \{k \in \mathbb{N}_0 \mid 0 \leq k \leq \min\{d(s), d(t)\}, k \bmod 2 = d(s) \bmod 2\}.$$

*Let $\mathcal{C}(G, k)$ denote the set of $(s,t)$-decompositions of $G$ in which there are $k$ $(s,t)$-simple paths. We define $\gamma(G, k) = |\mathcal{C}(G, k)|$.*

Using simple counting and parity arguments, we have the following observation.

**Observation 5** *For any legal graph $G = (V, E, s, t)$, $\mathcal{C}(G, k) = \emptyset$ for any $k \notin \kappa(G)$.*

**Lemma 6** *Let $G = (V, E, s, t)$ be an Eulerian graph with distinguished vertices $s, t$, let $k \in \kappa(G)$, and let $C$ be an $(s,t)$ decomposition of $G$ such that $C \in \mathcal{C}(G, k)$. Then the set of Euler tours of $G$ which are consistent with $C$ is a one-to-one correspondence with tuples of the form*

$$\left(\pi\{2,\ldots,k\}, \tau\{k+1,\ldots,\tfrac{k+d(s)}{2}\}, \sigma\{\tfrac{k+d(s)}{2}+1,\ldots,\tfrac{d(s)+d(t)}{2}\}, x, y, b \in \{1, -1\}^{\frac{d(s)+d(t)}{2}-k}\right), \quad (1)$$

*such that*

- $\pi, \tau, \sigma$ *are permutations on $k-1$, $\frac{d(s)-k}{2}$ and $\frac{d(t)-k}{2}$ elements respectively;*
- $x \in \mathbb{N}_0^{k/2}$ *is a sequence of $\frac{k}{2}$ non-negative integers which sums to $\frac{d(s)-k}{2}$;*
- $y \in \mathbb{N}_0^{k/2}$ *is a sequence of $\frac{k}{2}$ non-negative integers which sums to $\frac{d(t)-k}{2}$.*

**Proof:** Suppose that $C$ consists of $k$ $(s,t)$-simple paths, $k_s$ $(s,s)$-loops, and $k_t$ $(t,t)$-loops. Then since the paths of $C$ contain exactly one copy of each edge of $G$, it must be the case that $k_s = (d(s) - k)/2$ and $k_t = (d(t) - k)/2$. Moreover, $k$ must be an even number.

Let $p_1, \ldots, p_k$ be the $(s,t)$-simple paths in $C$; $p_{k+1}, \ldots, p_{(k+d(s))/2}$ be the $(s,s)$-loops in $C$, and $p_{(k+d(s))/2+1}, \ldots, p_{(d(s)+d(t))/2}$ be the $(t,t)$-loops in $C$. Assume wlog that $p_1$ is the path in $C$ whose $s$-adjacent edge has the lowest edge label, of any of the paths $p_1, \ldots, p_k$. We will assume wlog that $p_1$ is the initial path in every Euler tour, and that it is oriented from $s$ to $t$, thereby enforcing the rule that an Euler tour is not changed by a rotation or a reversal.

We will characterize all Euler tours consistent with $C$ by initially considering the order in which $p_2, \ldots, p_k$ appear in the tour after $p_1$ is traversed. This order is given by a permutation $\pi$ on $k-1$ elements. Note that all of the paths $p_{k+1}, \ldots, p_{(d(s)+d(t))/2}$ are $(s,s)$-loops or $(t,t)$-loops. Therefore if the ordering $p_{\pi(2)}, \ldots, p_{\pi(k)}$ is to be extendible to an Euler tour, then for every $2 \leq i \leq k$, we must direct the path $p_{\pi(i)}$ from $s$ to $t$ if and only if $i$ is odd (and from $t$ to $s$ if $i$ is even). Therefore $\pi$ determines both the order and the direction of all the paths $p_2, \ldots, p_k$ in the Euler tours for that $\pi$.

To complete the Euler tour, consider all ways of inserting the loops $p_{k+1}, \ldots, p_{(d(s)+d(t))/2}$ into the partial tour $p_1, p_{\pi(2)}, \ldots, p_{\pi(k)}$. In what follows, assume that $\pi(1) = 1$. Also identify $\pi(k+1)$ with $\pi(1)$ and $\pi(0)$ with $\pi(k)$. The $(s,s)$-loops can only be inserted into the intervals after some path $p_{\pi(2i)}$ and before $p_{\pi(2i+1)}$, for some $1 \leq i \leq \frac{k}{2}$. This implies that there are $k/2$ intervals where the $(d(s)-k)/2$ $(s,s)$-loops can be inserted. We may insert as many as $0$ or $(d(s)-k)/2$ of the $(s,s)$-loops into any of these $k/2$ positions, in any order. Moreover, each of the $(s,s)$-loops may be oriented in either of the two possible directions. These choices may be encoded in terms of

- A permutation $\tau$ on the paths $p_{k+1}, \ldots, p_{\frac{d(s)+k}{2}}$, specifying the order of insertion.

- A sequence $x \in \mathbb{N}_0^{k/2}$ of non-negative integers which sums to $\frac{d(s)-k}{2}$, where $x_i$ specifies the number of $(s,s)$-loops to be inserted between $p_{\pi(2i)}$ and $p_{\pi(2i+1)}$.

- A vector $b' \in \{-1, 1\}^{\frac{d(s)-k}{2}}$, specifying a direction for each of the $(s,s)$-loops.

Similarly, the $(t,t)$-loops can only be inserted into the intervals after some path $p_{\pi(2i-1)}$ and before $p_{\pi(2i)}$, for some $1 \leq i \leq \frac{k}{2}$. There are $(d(t)-k)/2$ $(t,t)$-loops, and $k/2$-positions where they may be inserted. These choices may be encoded in terms of

- A permutation $\sigma$ on the paths $p_{\frac{d(s)+k}{2}+1}, \ldots, p_{\frac{d(s)+d(t)}{2}}$, specifying the order of insertion.

- A sequence $y \in \mathbb{N}_0^{k/2}$ of non-negative integers which sums to $\frac{d(t)-k}{2}$, where $y_i$ specifies the number of $(t,t)$-loops to be inserted between $p_{\pi(2i-1)}$ and $p_{\pi(2i)}$.

- A vector $b'' \in \{-1, 1\}^{\frac{d(t)-k}{2}}$, specifying a direction for each of the $(t,t)$-loops.

We write $b = b'b''$ to obtain a sequence of length $(d(s)+d(t))/2 - k$ over $\{-1,1\}$.

Finally note that for any pair of Euler tours $T, T'$ that are both consistent with the decomposition $C$, either $T \sim T'$ under rotations and reversal (and therefore $T$ and $T'$ are the same tour), or else the tuple $(\pi(T), \tau(T), \sigma(T), x(T), y(T), b(T))$ induced by $T$ will differ from the tuple $(\pi(T'), \tau(T'), \sigma(T'), x(T'), y(T'), b(T'))$ for $T'$. This proves the one-to-one correspondence between the Euler tours with decomposition $C$ and the tuples $(\pi, \tau, \sigma, x, y, b)$. ∎

**Corollary 7** *Let $G = (V, E, s, t)$ be an Eulerian graph with distinguished vertices $s, t$. Then, for $\kappa(G)$ as given in Definition 3.2,*

$$|ET(G)| = \sum_{k \in \kappa(G)} \binom{k}{\frac{k}{2}} \frac{k}{4} 2^{(d(s)+d(t))/2-k} \left(\frac{d(s)}{2} - 1\right)! \left(\frac{d(t)}{2} - 1\right)! \gamma(G, k). \quad (2)$$

**Proof:** The number of $k$ for which $\mathcal{C}(G, k) \neq 0$ is finite. Any $C \in \mathcal{C}(G, k)$ must contain $(d(s) - k)/2$ $(s, s)$-loops and $(d(t) - k)/2$ $(t, t)$-loops in order to satisfy the conditions of an $(s, t)$-decomposition. Therefore we must have $k \leq d(s)$ and $k \leq d(t)$. Also, we require $k \bmod 2 = d(s) \bmod 2 = 0$.

By Observation 4, every $T \in ET(G)$ is consistent with a unique $(s, t)$-decomposition of $G$. We can therefore express $|ET(G)|$ as a sum over the set of $(s, t)$-decompositions, of the number of Euler tours consistent with that decomposition.

In Lemma 6 we proved that for any $k \in \kappa(G)$, the set of Euler tours consistent with any $(s, t)$-decomposition $C \in \mathcal{C}(G, k)$ is in one-to-one correspondence with the set of tuples described in (1). The *number* of these tuples depends on $k$, on $(d(s) - k)/2$ and on $(d(t) - k)/2$ but not on the specific decomposition itself. Therefore, we can express $|ET(G)|$ as a weighted sum of the $\gamma(G, k)$ values, where the weight against $\gamma(G, k)$ is the number of tuples for $k$ with respect to $G$. There are $(k - 1)!$ possible $\pi$ permutations, $((d(s) - k)/2)!$ possible $\tau$ permutations, and $((d(t) - k)/2)!$ possible $\sigma$ permutations. Counting all the possible $x$ vectors is equivalent to counting the number of sequences of $k/2$ non-negative numbers which sum to $(d(s) - k)/2$. This can be expressed as the number of ways of partitioning $(d(s) - k)/2$ into $k/2$ parts, which is $\binom{(d(s)-k)/2+(k/2)-1}{(k/2)-1} = \binom{(d(s)/2)-1}{(k/2)-1}$. Similarly the number of $y$ vectors satisfying the specified constraints is $\binom{(d(t)/2)-1}{(k/2)-1}$. The number of $b$-vectors is $2^{(d(s)+d(t))/2-k}$. Therefore the number of Euler tours which are consistent with $C$, for any $C \in \mathcal{C}(G, k)$, is exactly

$$(k-1)! \left(\frac{d(s)-k}{2}\right)! \left(\frac{d(t)-k}{2}\right)! \binom{\frac{d(s)}{2}-1}{\frac{k}{2}-1} \binom{\frac{d(t)}{2}-1}{\frac{k}{2}-1} 2^{\frac{d(s)+d(t)}{2}-k}$$

$$= (k-1)! \frac{(\frac{d(s)}{2}-1)!(\frac{d(t)}{2}-1)!}{(\frac{k}{2}-1)!(\frac{k}{2}-1)!} 2^{\frac{d(s)+d(t)}{2}-k}.$$

Then using the identity $\binom{k}{\frac{k}{2}} \frac{k}{4} = \frac{(k-1)!}{(((k/2)-1)!)^2}$, we get the desired expression for $|ET(G)|$. ∎

Corollary 7 gives an explicit formula to evaluate the number of Euler tours, given the value of $\gamma(G, k)$ for every feasible $k$. In the next section we will show how to recursively compute the $\gamma(G, k)$ values of a legal generalized series-parallel graph.

## 4 Recursive computation of $(s, t)$-decompositions

In Section 3, we have an exact formula for the number of Euler tours of a Eulerian graph, given the $\gamma(G, k)$ values for all feasible $k$. We now show how the $\gamma(G, k)$ values for a legal GSP graph $G$ can be recursively computed using the GSP structure of $G$. In Section 5 we apply these recurrences to exactly count and exactly-sample Euler tours.

### 4.1 Parallel combination

**Lemma 8** *Let $G_1 = (V_1, E_1, s, t)$ and $G_2 = (V_2, E_2, s', t')$ be two legal GSP graphs and consider the parallel composition $G_1 o_p G_2$, with source $s \sim s'$ and sink $t \sim t'$. Let $k \in \kappa(G_1 o_p G_2)$. The $(s, t)$-decompositions of $G_1 o_p G_2$ with $C \in \mathcal{C}(G_1 o_p G_2, k)$ are in one-to-one correspondence with pairs of the* form $(C_1, C_2)$, where $C_1 \in \mathcal{C}(G_1, k_1)$, $C_2 \in \mathcal{C}(G_2, k - k_1)$ for some $k_1$ such that $k_1 \in \kappa(G_1)$ and $k - k_1 \in \kappa(G_2)$.

**Proof:** Let $C \in \mathcal{C}(G_1 o_p G_2, k)$. By construction of $G_1 o_p G_2$, we know that any $(s, t)$-simple path, $(s, s)$-loop or $(t, t)$-loop of $G_1 o_p G_2$ either lies entirely in $G_1$ or entirely in $G_2$. Define $C_1$ to be the set of paths and loops of $C$ which lie entirely in $G_1$. Define $C_2$ to be the set of paths and loops (with $s'$ substituted for $s$) which lie entirely in $G_2$. The pair $(C_1, C_2)$ is unique by this definition. Then $C_1$ is a $(s, t)$-decomposition for $G_1$ for some $k_1 \in \kappa(G_1)$, $k_1 \leq k$, and $C_2$ is an $(s', t')$-decomposition with $k - k_1$ $(s', t')$-simple paths.

On the other hand, consider any pair $(C_1, C_2)$ where $C_1 \in \mathcal{C}(G_1, k_1)$ and $C_2 \in \mathcal{C}(C_2, k - k_1)$, such that $k_1 \in \kappa(G_1)$ and $k - k_1 \in \kappa(G_2)$. Then $C_1 \cup \widehat{C_2} \in \mathcal{C}(G_1 o_p G_2, k_1 + k_2)$, where $\widehat{C_2}$ is a copy of $C_2$ with $s', t'$ replaced by $s, t$ respectively.

These two parts prove the one-to-one correspondence. ∎

**Corollary 9** *Let $G_1 = (V_1, E_1, s, t)$ and $G_2 = (V_2, E_2, s', t')$ be two legal GSP graphs and consider the parallel composition $G_1 o_p G_2$, with source $s \sim s'$ and sink $t \sim t'$. Let $k \in \kappa(G_1 o_p G_2)$. Then the number of $(s, t)$-decompositions of $G_1 o_p G_2$ which contain $k$ $(s, t)$-simple paths is*

$$\gamma(G_1 o_p G_2, k) = \sum_{k_1 \in \kappa(G_1), k - k_1 \in \kappa(G_2)} \gamma(G_1, k_1) * \gamma(G_2, k - k_1).$$

## 4.2 Series combination

The recursive relationship between the set of $(s, t')$-decompositions of $G_1 o_s G_2$ and the $(s, t)$-decompositions of $G_1$ and $(s', t')$-decompositions of $G_2$ is more involved than for the parallel case. We will see in Lemma 10 that the decompositions which are used to form $(s, t')$-decompositions of $\mathcal{C}(G_1 o_s G_2, k)$ are the elements of $\mathcal{C}(G_1, k_1)$ for $k_1 \geq k$ and of $\mathcal{C}(G_2, k_2)$ for $k_2 \geq k$.

**Definition 4.1** *Suppose we are given two series parallel graphs $G_1 = (V_1, E_1, s, t)$ and $G_2 = (V_2, E_2, s', t')$ such that $d_{G_1}(s) \bmod 2 = d_{G_2}(s') \bmod 2$. Suppose $k_1 \in \kappa(G_1)$ and $k_2 \in \kappa(G_2)$. We define $D(k_1, k_2)$ (with respect to $G_1, G_2$) to be*

$$D(k_1, k_2) = d_{G_1}(t) + d_{G_2}(s') - k_1 - k_2.$$

**Lemma 10** *Let $G_1 = (V_1, E_1, s, t)$ and $G_2 = (V_2, E_2, s', t')$ be two legal GSP graphs such that $d_{G_1}(s) \bmod 2 = d_{G_2}(s') \bmod 2$. Consider the series composition $G_1 o_s G_2$ of the two graphs, with source $s$ and sink $t'$. Then for any $k \in \kappa(G_1 o_s G_2)$, there is a one-to-many correspondence between the elements of $\mathcal{C}(G_1 o_s G_2, k)$ and between tuples of the form*

$$\bigcup_{\substack{k_1 \in \kappa(G_1) \\ k_1 \geq k}} \bigcup_{\substack{k_2 \in \kappa(G_2) \\ k_2 \geq k}} \left( C_1, C_2, \pi_1\{1, \ldots, k_1\}, \pi_2\{1, \ldots, k_2\}, x \in \mathbb{N}_0^{(k_1+k_2)/2}, \sigma\{1, \ldots, \tfrac{D(k_1,k_2)}{2}\}, \{1, -1\}^{D(k_1,k_2)/2} \right),$$

*such that*

- $C_1$ *is an $(s, t)$-decomposition from $\mathcal{C}(G_1, k_1)$ and $C_2$ is an $(s', t')$-decomposition from $\mathcal{C}(G_2, k_2)$.*

- $\pi_1, \pi_2, \sigma$ *are permutations on $k_1$ elements, $k_2$ elements and $\tfrac{D(k_1,k_2)}{2}$ elements respectively.*

- $x$ *is a non-negative integer vector of length $(k_1 + k_2)/2$ which sums to $D(k_1, k_2)/2$.*

*The factor in the one-to-many correspondence is $\Phi_s(k, k_1, k_2) =_{def} k! \tfrac{k_1-k}{2}! \tfrac{k_2-k}{2}! 2^{(k_1+k_2)/2-k}$.*

**Proof:** First consider an arbitrary $(s, t')$-decomposition $C$ of $G_1 o_s G_2$ with $C \in \mathcal{C}(G_1 o_s G_2, k)$. Recall that $t \sim s'$ is the point where $G_1$ was joined to $G_2$. For every path $p \in C$, let $p_1, \ldots, p_{\ell(p)}$ be the set of segments of $p$ such that for each $p_i$, the endpoints, but none of the intermediate points of $p_i$ lie in $\{s, t', t \sim s'\}$, and such that each segment is directed such that the initial edge has the lower edge label. Then for every path $p \in C$ and every $1 \leq i \leq \ell(p)$, we know that $p_i$ either lies entirely in $G_1$ or entirely in $G_2$. We define

$$\begin{aligned} C_1 &= \cup_{p \in C} \{p_i : 1 \leq i \leq \ell(p), p_i \in G_1\} \\ C_2 &= \cup_{p \in C} \{p_i : 1 \leq i \leq \ell(p), p_i \in G_2\}. \end{aligned}$$

$C_1$ is then an $(s, t)$-decomposition of $G_1$ and $C_2$ is an $(s', t')$-decomposition of $G_2$. Suppose that $C$ was an element of $\mathcal{C}(G_1 o_s G_2, k)$ for $k \in \kappa(G_1 o_s G_2)$. Then each of the $k$ $(s, t')$-simple paths of $C$ will contribute one $(s, t)$-simple path to $C_1$ and one $(s', t')$-simple path to $C_2$. This is because every path between $s$ and $t'$ in $G_1 o_s G_2$ must pass through $t \sim s'$. Hence if $C_1 \in \mathcal{C}(G_1, k_1)$ and $C_2 \in \mathcal{C}(G_2, k_2)$, we are guaranteed that $k \leq k_1$ and $k \leq k_2$. By our rule for directing the $p_i$ segments, the pair $(C_1, C_2)$ is uniquely defined for any given $C$.

Now suppose we are given $C_1 \in \mathcal{C}(G_1, k_1)$, $C_2 \in \mathcal{C}(G_2, k_2)$ for $k_1 \in \kappa(G_1), k_2 \in \kappa(G_2)$. We will characterize the $(s, t')$-decompositions of $G_1 o_s G_2$ which can be constructed from $C_1$ and $C_2$. First of all note that by the argument of the previous paragraph we can only use $(C_1, C_2)$ to construct decompositions which contain at most $\min\{k_1, k_2\}$ paths with both $s$ and $t'$ as endpoints. Suppose we are considering such a $k \in \kappa(G_1 o_s G_2)$, $k \leq \min\{k_1, k_2\}$.

In $C_1$ we have $k_1$ $(s, t)$-simple paths, $(d_{G_1}(s) - k_1)/2$ $(s, s)$-loops, and $(d_{G_1}(t) - k_1)/2$ $(t, t)$-loops. In $C_2$ we have $k_2$ $(s', t')$-simple paths, $(d_{G_2}(s') - k_2)/2$ $(s', s')$-loops, and $(d_{G_2}(t') - k_2)/2$ $(t', t')$-loops. For $C$, we need to construct $k$ $(s, t')$-simple paths, $(d_{G_1}(s) - k)/2$ $(s, s)$-loops, and $(d_{G_2}(t') - k)/2$ $(t', t')$-loops. We make the following observations:

$(s, t')$**-simple paths:** By construction of $G_1 o_s G_2$, every path between $s$ and $t'$ in $G_1 o_s G_2$ must pass through $t \sim s'$. Therefore, every $(s, t')$-simple path in $C$ must be built using exactly one of the $k_1$ $(s, t)$-simple paths of $C_1$, and exactly one of the $k_2$ $(s', t')$-simple paths of $C_2$.

$(s, s)$**-loops:** Each of the $(s, s)$-loops of $C_1$ must become an $(s, s)$-loop in $C$. The extra number of $(s, s)$-loops that we will need to add to $C$ is $(d_{G_1}(s) - k - (d_{G_1}(s) - k_1))/2 = (k_1 - k)/2$.

$(t', t')$**-loops:** Each of the $(t', t')$-loops of $C_2$ must become a $(t', t')$-loop in $C$. The extra number of $(t', t')$-loops that we will need to add to $C$ is $(d_{G_2}(t') - k - (d_{G_2}(t') - k_2))/2 = (k_2 - k)/2$.

Observe that after $k$ paths have been chosen from the $(s, t)$-simple paths of $C_1$ and from the $(s', t')$-simple paths of $C_2$, the paths and loops of $C_1 \cup C_2$ which have not been allocated any role in $C$ are as follows:

- $k_1 - k$ remaining $(s, t)$-simple paths from $C_1$,
- $k_2 - k$ remaining $(s', t')$-simple paths from $C_2$,
- all of the $(d_{G_1}(t) - k_1)/2$ $(t, t)$-loops from $C_1$,
- all of the $(d_{G_2}(s') - k_2)/2$ $(s', s')$-loops from $C_2$.

Neither the $(t, t)$-loops from $C_1$ nor the $(s', s')$-loops from $C_2$ contain either of the distinguished vertices $s$ and $t'$ of $G_1 o_s G_2$; therefore they play no significant role in constructing the extra $(s, s)$-loops nor the extra $(t', t')$-loops needed for $C$. The $(k_1 - k)$ remaining $(s, t)$-simple paths from $C_1$ must therefore be used to construct the extra $(k_1 - k)/2$ $(s, s)$-loops for $C$. Hence, we only need to construct a pairing of these remaining $(s, t)$-simple paths that were not used for the $(s, t')$-simple paths of $C$. Similarly, in any $C$ which is built from the paths in $C_1 \cup C_2$, we must use the remaining $(k_2 - k)$ $(s', t')$-simple paths to construct the extra $(k_2 - k)/2$ $(t', t')$-loops needed for $C$. Again, any pairing of these $(k_2 - k)$-paths is sufficient.

We now observe that the choice-and-pairing of the $k$ $(s, t)$-simple paths from $C_1$ and the $k$ $(s', t')$-simple paths from $C_2$ to form the $(s, t')$-simple paths of $C$ can actually be combined with the pairing of the remaining $(k_1 - k)$ $(s, t)$-simple paths, and also the pairing of the remaining $(k_2 - k)$ $(s', t')$-simple paths, as follows: we permute all of the $(s, t)$-simple paths of $C_1$ (a permutation $\pi_1$ of length $k_1$) and also permute all of the $(s', t')$-simple paths of $C_2$ (a permutation of length $k_2$). Then we construct the following pairings:

- $(s, t')$-pairing: We pair the $\pi_1(i)$-th $(s, t)$-simple path of $C_1$ with the $\pi_2(i)$-th $(s', t')$-simple path of $C_2$, for every $1 \leq i \leq k$.

- $(s, s)$-pairings: We already have $(d_{G_1}(s) - k)/2$ $(s, s)$-loops from $C_1$. The extra $(k_1 - k)/2$ $(s, s)$-loops are constructed by pairing $\pi_1(i)$ with $\pi_1(i + 1)$ for $k < i < k_1$, $i \bmod 2 = (k + 1) \bmod 2$.

- $(t', t')$-pairings: We already have $(d_{G_2}(t') - k)/2$ $(t', t')$-loops from $C_2$. The extra $(k_2 - k)/2$ $(t', t')$-loops are constructed by pairing $\pi_2(i)$ with $\pi_2(i + 1)$ for $k < i < k_2$, $i \bmod 2 = (k + 1) \bmod 2$.

Observe that in this model, the number of ways of coming up with the same set of pairings is $k!(\frac{k_1-k}{2})!(\frac{k_2-k}{2})! 2^{(k_1+k_2)/2-k}$. Observe that for all these pairings, our requirement to order the paths of $C$ in terms of lower-edge first implies that the choice of the pairing determines the relative order of the pairs, for every pairing we have constructed.

Finally, the $(t, t)$-loops of $C_1$ and $(s', s')$-loops of $C_2$ must be included in our decomposition $C$. There are $D(k_1, k_2)/2$ of these in total. These paths may be inserted at any point where $t$ appears in the partial paths we have constructed so far. There are exactly $(k_1 + k_2)/2$ occurrences of $t \sim s'$ in the $(s, t')$, $(s, s)$ and $(t, t)$-pairings we have constructed at this point. We may insert as many as 0 or all $D(k_1, k_2)/2$ of the $(t, t)$ and $(s', s')$ loops into any individual position - the particular partition chosen is encoded as the $x$ vector. Each of the $D(k_1, k_2)/2$ items is different, so once the partitioning has been determined, there are $\frac{D(k_1,k_2)}{2}!$ ways of ordering the $(t, t)$- and $(s', s')$-loops for insertion - encoded by the $\sigma$ permutation. Finally, no $(t, t)$- or $(s', s')$-loop $q$ will ever be inserted so that it is adjacent to an endpoint of any path of $C$. Therefore for every such $q$, the 2 directions of inserting $q$ ($q$ and $rev(q)$) result in a different $C$-decomposition. ∎

**Corollary 11** *Let $G_1 = (V_1, E_1, s, t)$ and $G_2 = (V_2, E_2, s', t')$ be two legal GSP graphs such that $d_{G_1}(s) \bmod 2 = d_{G_2}(s') \bmod 2$. Consider the* series composition $G_1 o_s G_2$ *of the two graphs, with source $s$ and sink $t'$. Then for any $k \in \kappa(G_1 o_s G_2)$, the number of $(s, t)$-decompositions of $G_1 o_s G_2$ with $k$ $\{s, t\}$-simple paths is $\gamma(G_1 o_s G_2, k)$*

$$= \sum_{\substack{k_1 \in \kappa(G_1) \\ k_1 \geq k}} \sum_{\substack{k_2 \in \kappa(G_2) \\ k_2 \geq k}} \gamma(G_1, k_1) * \gamma(G_2, k_2) * \frac{k_1! k_2!}{k! \frac{k_1-k}{2}! \frac{k_2-k}{2}!} * \frac{\left(\frac{d_{G_1}(t)+d_{G_2}(s')}{2} - 1\right)!}{\left(\frac{k_1+k_2}{2} - 1\right)!} * \frac{2^{(d_{G_1}(t)+d_{G_2}(s'))/2}}{2^{k_1+k_2-k}}.$$

**Proof:** We use Lemma 10. For any particular $k_1, k_2$, there are $k_1!$ $\pi_1$ permutations, $k_2!$ $\pi_2$ permutations, and $(D(k_1, k_2)/2)!$ $\sigma$ permutations. The number of $x$ vectors is the same as the number of ways of partitioning $D(k_1, k_2)$ items into $(k_1 + k_2)/2$ parts, which is

$$\binom{(k_1+k_2)/2 + D(k_1,k_2)/2 - 1}{(k_1+k_2)/2 - 1} = \frac{(\frac{d_{G_1}(t)+d_{G_2}(s')}{2} - 1)!}{(\frac{k_1+k_2}{2} - 1)! \frac{D(k_1,k_2)}{2}!}.$$

The number of tuples for a particular $C_1, C_2$ with $C_1 \in \mathcal{C}(G_1, k_1)$ and $C_2 \in \mathcal{C}(G_2, k_2)$, is therefore

$$k_1! k_2! \frac{(\frac{d_{G_1}(t)+d_{G_2}(s')}{2} - 1)!}{(\frac{k_1+k_2}{2} - 1)! \frac{D(k_1,k_2)}{2}!} \frac{D(k_1,k_2)}{2}! 2^{D(k_1,k_2)/2}.$$

Dividing by $k!\frac{k_1-k}{2}!\frac{k_2-k}{2}!2^{(k_1+k_2)/2-k}$, and cancelling some terms, we find that the number of $(s,t')$-decompositions of $G_1o_sG_2$ with $k$ $(s,t')$-simple paths that can be constructed from $C_1, C_2$ is

$$\frac{k_1!k_2!}{k!(\frac{k_1-k}{2})!(\frac{k_2-k}{2})!} \frac{(\frac{d_{G_1}(t)+d_{G_2}(s')}{2}-1)!}{(\frac{k_1+k_2}{2}-1)!} \frac{2^{(d_{G_1}(t)+d_{G_2}(s'))/2}}{2^{k_1+k_2-k}}.$$

∎

### 4.3 Dangling combination

The dangling combination changes the degree of the source and leaves the degree of the sink unaffected in a GSP graph. In terms of the $(s,t)$-decomposition, the dangling operation adds $(s,s)$-loops at the source $s$.

**Lemma 12** *Let $G_1 = (V_1, E_1, s, t)$ be a legal GSP graph and $G_2 = (V_2, E_2, s', t')$ be an Eulerian GSP graph and consider the dangling composition $G_1o_dG_2$, with source $s \sim s'$ and sink $t$. For every $k \in \kappa(G_1o_dG_2) \setminus \kappa(G_1)$, $\mathcal{C}(G_1o_dG_2, k) = \emptyset$. For every $k \in \kappa(G_1)$, the elements of $\mathcal{C}(G_1o_dG_2, k)$ are in one-to-many correspondence with tuples of the form*

$$\bigcup_{k_2 \in \kappa(G_2)} \left(C_1, C_2, \pi_2\{1, \ldots, k_2\}, x \in \mathbb{N}_0^{k_2/2}, \sigma\{1, \ldots, \frac{d_{G_2}(t')-k_2}{2}\}, \{1,-1\}^{(d_{G_2}(t')-k_2)/2}\right),$$

*such that*

- $C_1$ *is an $(s,t)$-decomposition from $\mathcal{C}(G_1, k)$ and $C_2$ is an $(s',t')$-decomposition from $\mathcal{C}(G_2, k_2)$ where $k_2 \in \kappa(G_2)$.*

- $\pi_2, \sigma$ *are permutations of length $k_2$ and $(d_{G_2}(t') - k_2)/2$ respectively.*

- $x$ *is a non-negative integer vector of length $k_2/2$ which sums to $(d_{G_2}(t') - k_2)/2$.*

*The factor in the one-to-many correspondence is $\Phi_d(k_2) =_{def} \frac{k_2}{2}! \, 2^{k_2/2}$.*

**Proof:** First observe that for any $k$, and any $C \in \mathcal{C}(G_1o_dG_2, k)$, $C$ induces an $(s,t)$-decomposition $C'$ on $G_1$ with exactly $k$ $(s,t)$-simple paths. Therefore we must have $\mathcal{C}(G_1o_dG_2, k) = \emptyset$ for every $k \notin \kappa(G_1)$.

Now, for any $k \in \kappa(G_1)$ consider an arbitrary $(s,t)$-decomposition $C$ of $G_1o_dG_2$ with $C \in \mathcal{C}(G_1o_dG_2, k)$. Recall that $s \sim s'$ is the point where $G_1$ is joined to $G_2$. For every path $p \in C$, let $p_1, \ldots, p_{\ell(p)}$ be the set of segments of $p$ such that for each $p_i$, the endpoints, but none of the intermediate points of $p_i$ lie in $\{s(\sim s'), t, t'\}$, and such that each segment is directed such that the initial edge has the lower edge label. Then for every path $p \in C$ and every $1 \leq i \leq \ell(p)$, we know that $p_i$ either lies entirely in $G_1$ or entirely in $G_2$. We define

$$C_1 = \cup_{p \in C}\{p_i : 1 \leq i \leq \ell(p), p_i \in G_1\}$$
$$C_2 = \cup_{p \in C}\{p_i : 1 \leq i \leq \ell(p), p_i \in G_2\}.$$

$C_1$ is then an $(s,t)$-decomposition of $G_1$ and $C_2$ is an $(s',t')$-decomposition of $G_2$. Suppose that $C$ is an element of $\mathcal{C}(G_1o_dG_2, k)$ for $k \in \kappa(G_1o_dG_2)$. Then a path $p \in C$ is an $(s,t)$-simple path of $C$ if and only if it is an $(s,t)$-simple path of $C_1$. This is because every $(s,t)$-simple path of $C$ has to lie entirely in $G_1$ and hence is an $(s,t)$ simple path in $C_1$. This, in turn, implies that $C_1 \in \mathcal{C}(G_1, k)$ and $C_2 \in \mathcal{C}(G_2, k_2)$ where $k_2$ is independent of $k$ and $k_2 \in \kappa(G_2)$. By our rule for directing the $p_i$ segments, the pair $(C_1, C_2)$ is uniquely defined for any given $C$.

Now suppose we are given $C_1 \in \mathcal{C}(G_1, k_1)$ and $C_2 \in \mathcal{C}(G_2, k_2)$ where $k_1 \in \kappa(G_1), k_2 \in \kappa(G_2)$. We will characterize the $(s,t)$-decompositions of $G_1o_dG_2$ which can be constructed from $C_1$ and $C_2$. First of all note that by the argument of the previous paragraph, an $(s,t)$-decomposition of $G_1 o_d G_2$ constructed using $(C_1, C_2)$ will have exactly $k_1$ $(s,t)$-simple paths. In other words, if $C \in \mathcal{C}(G_1 o_d G_2, k)$ is constructed using $(C_1, C_2)$, then $C_1 \in \mathcal{C}(G_1, k_1)$ where $k_1 \in \kappa(G_1)$ and $k_1 = k$. We also note that a path $p \in C$ is a $(t,t)$-loop in $C$ if and only if $p$ is an $(t,t)$-loop in $C_1$. This is because the vertex $t$ lies in $V(G_1) \setminus V(G_2)$.

In $C_1$ we have $k_1$ $(s,t)$-simple paths, $(d_{G_1}(t) - k_1)/2$ $(t,t)$-loops, and $(d_{G_1}(s) - k_1)/2$ $(s,s)$-loops. In $C_2$ we have $k_2$ $(s',t')$-simple paths, $(d_{G_2}(t') - k_2)/2$ $(t',t')$-loops, and $(d_{G_2}(s') - k_2)/2$ $(s',s')$-loops. For $C$, we need to construct $k(= k_1)$ $(s,t)$-simple paths, $(d_{G_1}(t)-k)/2$ $(t,t)$-loops, and $(d_{G_1}(s)+d_{G_2}(s')-k)/2$ $(s,s)$-loops. We make the following observation:

$(s,s)$-**loops:** Every $(s,s)$-loop of $C_1$ and every $(s',s')$-loop of $C_2$ must become an $(s,s)$-loop in $C$. The extra number of $(s,s)$-loops that we will need to add to $C$ is $(d_{G_1}(s) + d_{G_2}(s') - k)/2 - (d_{G_1}(s) - k)/2 - (d_{G_2}(s') - k_2)/2 = k_2/2$.

Observe that the $(s,t)$-simple paths and $(t,t)$-loops of $C_1$ have become the $(s,t)$-simple paths and $(t,t)$-loops of $C$ and the $(s,s)$-loops of $C_1$ and $C_2$ have become the $(s,s)$-loops of $C$. The paths and loops of $C_1 \cup C_2$ which have not been allocated any role in $C$ are as follows:

- all of the $k_2$ $(s',t')$-simple paths from $C_2$,
- all of the $(d_{G_2}(t') - k_2)/2$ $(t',t')$-loops from $C_2$,

All of the $k_2$ $(s',t')$-simple paths and the $(d_{G_2}(t') - k_2)/2$ $(t',t')$-loops from $C_2$ would be combined to form $k_2/2$ $(s,s)$-loops in $C$ ( extra $(s',s')$-loops in $C_2$). As the $(t',t')$-loops do not contain the vertex $s'$, we ignore them for the time being. To obtain the $k_2/2$ $(s',s')$-loops, we pair up the $k_2$ $(s',t')$-simple paths in the following fashion. We permute all the $(s',t')$-simple paths (a permutation $\pi_2$ of length $k_2$) and construct the following pairing:

$(s',s')$-pairing: The extra $(s',s')$ loops are constructed by pairing the $\pi_2(i)$-th $(s',t')$-simple path of $C_2$ with the $\pi_2(i+1)$-th $(s',t')$-simple path of $C_2$, where $1 \leq i \leq k_2$ and $i \equiv 1 \pmod 2$.

Observe that in this model, the number of ways of coming up with the same set of pairings is $\frac{k_2}{2}! \, 2^{k_2/2}$. Observe that for all these pairings, our requirement to order the paths of $C$ in terms of lower-edge first implies that the choice of the pairing determines the relative order of the pairs, for every pairing we have constructed.

Finally, the $(t',t')$-loops of $C_2$ must be included in our decomposition $C$. There are $(d_{G_2}(t') - k_2)/2$ of these in total. These loops may be inserted at any point where $t'$ appears in the partial $(s',s')$-loops we have constructed so far. There are exactly $k_2/2$ occurrences of $t'$ in the extra $k_2/2$ $(s',s')$-pairings we have constructed at this point. We may insert as many as 0 or all $(d_{G_2}(t') - k_2)/2$ of the $(t',t')$ loops into any individual position - the particular partition chosen is encoded as the $x$ vector. Each of the $(d_{G_2}(t') - k_2)/2$ items is different, so once the partitioning has been determined, there are $\frac{d_{G_2}(t')-k_2}{2}!$ ways of ordering the $(t',t')$-loops for insertion - encoded by the $\sigma$ permutation. Finally, no $(t',t')$-loop $q$ will ever be inserted so that it is adjacent to an endpoint of any path of $C$. Therefore for every such $q$, the 2 directions of inserting $q$ ($q$ and $rev(q)$) result in a different $C$-decomposition. ∎

**Corollary 13** *Let $G_1 = (V_1, E_1, s, t)$ be a legal GSP graph and $G_2 = (V_2, E_2, s', t')$ be an Eulerian GSP graph. Consider the dangling composition $G_1 o_d G_2$ of the two graphs, with source $s$ and sink $t$. Then for any $k \in \kappa(G_1 o_d G_2) = \kappa(G_1)$, the number of $(s,t)$-decompositions of $G_1 o_d G_2$ with $k$ $\{s,t\}$-simple paths is $\gamma(G_1 o_d G_2, k)$*

$$= \gamma(G_1, k) * \sum_{k_2 \in \kappa(G_2)} \gamma(G_2, k_2) * \frac{k_2!}{\frac{k_2}{2}!(\frac{k_2}{2}-1)!} * \left(\frac{d_{G_2}(t')}{2} - 1\right)! * 2^{d_{G_2}(t')/2 - k_2}.$$

**Proof:** We use Lemma 12. For any given $k_2$ there are $k_2!$ $\pi_2$ permutations and $((d_{G_2}(t') - k_2)/2)!$ $\sigma$ permutations. The number of $x$ vectors is the same as the number of ways of partitioning $(d_{G_2}(t') - k_2)/2$ items into $k_2/2$ parts, which is

$$\binom{k_2/2 + (d_{G_2}(t') - k_2)/2 - 1}{(k_2/2) - 1} = \frac{(\frac{d_{G_2}(t')}{2} - 1)!}{(\frac{k_2}{2} - 1)! \left(\frac{d_{G_2}(t') - k_2}{2}\right)!}.$$

The number of tuples for a particular $C_1, C_2$ with $C_1 \in \mathcal{C}(G_1, k)$ and $C_2 \in \mathcal{C}(G_2, k_2)$, is therefore

$$1 * k_2! \left(\frac{d_{G_2}(t') - k_2}{2}\right)! \frac{(\frac{d_{G_2}(t')}{2} - 1)!}{(\frac{k_2}{2} - 1)! \left(\frac{d_{G_2}(t') - k_2}{2}\right)!} 2^{(d_{G_2}(t') - k_2)/2}.$$

Dividing by $\frac{k_2}{2}! \, 2^{k_2/2}$, and cancelling some terms, we find that the number of $(s,t)$-decompositions of $G_1 o_d G_2$ with $k$ $(s,t)$-simple paths that can be constructed from $C_1, C_2$ is

$$\frac{k_2!}{\frac{k_2}{2}!(\frac{k_2}{2} - 1)!} * \left(\frac{d_{G_2}(t')}{2} - 1\right)! * 2^{d_{G_2}(t')/2 - k_2}.$$

∎

## 5 Algorithms

Our counting and sampling algorithms will use a compact binary tree representation $T(G)$ for the GSP graph $G$, as in Figure 2. This representation can be computed in polynomial-time (see [6]), and is of size linear in the number of edges of $G$.

### 5.1 Counting Euler tours

Consider a binary tree representation $T$ of a given Eulerian GSP graph $G = (V, E, s, t)$. We assume that every vertex $u \in V(T)$ represents a subgraph $G_u$ of the graph $G$, obtained by applications of $o_s, o_p,$ or $o_d$ to the graphs within the subtree at rooted at $u$. As noted in Observation 2, such component graph $G_u$ in a hierarchical decomposition of $G$ will satisfy the property of being a legal GSP graph. For each vertex $u$ in $T$ let $s_u$ denote the source of $G_u$ and $t_u$ the sink of $G_u$.

For each such $u \in V(T)$, we compute the values of $\gamma(G_u, k)$ for $k \in \kappa(G_u)$. To do this, for each $u \in V(T)$, we define variables $d_u(s_u), d_u(t_u)$ and the integer array $\gamma_u(k)$, for $1 \leq k \leq m$ (where $m = |E|$) with each vertex $u$ of the tree. The value $d_u(s_u)$ will be used to store the degree of the source $s_u$ in $G_u$ and the value $d_u(t_u)$ will be used to store the degree of the sink $t_u$ in $G_u$. For every $k \in \kappa(G_u)$, $\gamma_u(k)$ will be used to store the value of $\gamma(G_u, k)$. All values for the internal nodes are initialized to 0. We compute the values for $d_u(s_u), d_u(t_u)$ and $\gamma_u(k)$ for all nodes $u \in V(T)$ in a bottom-up fashion.

Each leaf of $T(G)$ corresponds to a graph consisting of a single edge of the graph $G$, ie, to a $B$ operation. Hence we set $d_u(s_u) = 1$ and $d_u(t_u) = 1$ for every leaf $u$. For a graph $G_u$ consisting of a single edge, we have $\kappa(G_u) = \{1\}$. Therefore we set $\gamma_u(1) = 1$ and $\gamma_u(k) = 0$ for $k \neq 1$.

Alternatively, if $u$ is not a leaf of $T$, then $G_u$ was obtained from the two smaller legal GSP graphs corresponding to the child nodes of $u$ via the $o_s$ or the $o_p$ or the $o_d$ operation. Let $\ell$ and $r$ denote the left and right children of $u$. We set the values of $d_u(s_u)$ and $d_u(t_u)$ as follows:

$$\begin{aligned}
\text{Series operation} &: s_u = s_\ell, d_u(s_u) = d_\ell(s_\ell), \quad t_u = t_r, d_u(t_u) = d_r(t_r) \\
\text{Parallel operation} &: s_u = s_\ell = s_r, d_u(s_u) = d_\ell(s_\ell) + d_r(s_r), \\
& \quad t_u = t_\ell = t_r, d_u(t_u) = d_\ell(t_\ell) + d_r(t_r) \\
\text{Dangling operation} &: s_u = s_\ell = s_r, d_u(s_u) = d_\ell(s_\ell) + d_r(s_r), \\
& \quad t_u = t_\ell, d_u(t_u) = d_\ell(t_\ell)
\end{aligned}$$

In the case of $o_p$, we use the equation of Corollary 9 to compute the value of $\gamma_u(k)$ from the $\gamma_\ell(\cdot)$ and $\gamma_r(\cdot)$ values, for every $k \in \kappa(G_u)$. In the case of $o_s$, we use the equation of Corollary 11 to compute the value of $\gamma_u(k)$ from the $\gamma_\ell(\cdot)$ and $\gamma_r(\cdot)$ values. In the case of $o_d$, we use the equation of Corollary 13 to compute the value of $\gamma_u(k)$ from the $\gamma_\ell(\cdot)$ and $\gamma_r(\cdot)$ values.

Finally, once we have the values of $\gamma(G, k)$ for $k \in \kappa(G)$ for the given GSP graph $G$, we can use equation (2) to compute $|ET(G)|$.

## 5.2 Time and space complexity of counting

Given an Eulerian GSP, $G$, what is the time complexity of computing $|ET(G)|$? We first assume that we already have a binary tree decomposition that describes how to construct $G$ (as described in Section 2.1) using the standard operations for building GSPs. Note, first, that if $|E(G)| = m$, then the binary tree has $m$ leaves and $m - 1$ internal vertices, as each operation in the construction of $G$ combines two legal subgraphs to make one new connected (legal) subgraph.

Corollary 7 tells us how to find $|ET(G)|$, given the values of $\gamma(G, k)$ for all $k \in \kappa(G)$. Let $\Delta = $ max degree of $G$. From Corollary 7, we see that the number of terms in the sum for computing $|ET(G)|$ is at most $\Delta/2$ (as $k$ must agree with the parity of the degree of the source $s$). So we need to know how much time is required to compute the $\gamma(G, k)$ values for any fixed $k \in \kappa(G)$. We consider each operation in turn, and examine the time needed to compute the values $\gamma(G_1 o_x G_2, k)$ for an operation $o_x \in \{o_p, o_s, o_d\}$, given all the corresponding values for the two graphs $G_1$ and $G_2$.

**Parallel combination:** For a fixed $k \in \kappa(G_1 o_p G_2)$, from Corollary 9 we see there are $\min\{|\kappa(G_1)|, |\kappa(G_2)|\}$ terms in the sum to find $\gamma(G_1 o_p G_2, k)$, given the values for each of $G_1$ and $G_2$. Therefore, computing $\gamma(G_1 o_p G_2, k)$ takes $\mathcal{O}(\Delta)$ operations. (Note: Here we are assuming that we before beginning our computations to find $|ET(G)|$, we first perform a "pre-processing step" by computing, and storing, values $k!$ and $2^k$ for $0 \leq k \leq \Delta$. We can do this "pre-processing" in time $\mathcal{O}(\Delta)$, using $\mathcal{O}(\Delta^2 \log \Delta)$ and $\mathcal{O}(\Delta^2)$, respectively, bits of space for the $k!$ and $2^k$ values.)

**Series combination:** Corollary 11 tells us there are at most $|\kappa(G_1)| \cdot |\kappa(G_2)| \in \mathcal{O}(\Delta^2)$ terms in the (double) summation for finding $\gamma(G_1 o_s G_2), k$ for a fixed $k \in \kappa(G_1 o_s G_2)$.

**Dangling combination:** Finally, from Corollary 13 we see there are $|\kappa(G_2)| \in \mathcal{O}(\Delta)$ terms in the summation for finding $\gamma(G_1 o_d G_2, k)$ for a fixed $k \in \kappa(G_1 o_d G_2)$.

Overall, we see that computing any value $\gamma(H, k)$, for any legal subgraph $H$ of $G$ (corresponding to an internal vertex of the binary tree decomposition), and any fixed $k \in \kappa(H)$, takes time $\mathcal{O}(\Delta^2)$. Since there are at most $\mathcal{O}(\Delta)$ values of $k \in \kappa(H)$, and $m - 1$ internal vertices in the binary tree decomposition, we can compute all $\gamma(H, k)$ values with $\mathcal{O}(m\Delta^3)$ operations. Finally, since there are $\mathcal{O}(\Delta)$ terms in the sum to find $|ET(G)|$, we see we can find the number of Euler tours of a GSP in time $\mathcal{O}(m\Delta^3)$.

Taking a crude upper bound of $\mathcal{O}\left(\binom{\Delta}{2}\binom{\Delta-2}{2}\cdots\binom{2}{2}\right) \in \mathcal{O}(\Delta^\Delta)$ for the number of Euler tours of an Eulerian graph with max degree $\Delta$, we see that we need at most $\mathcal{O}(\Delta \log \Delta)$ bits to store one $\gamma(H,k)$ value. Using, again, that there are $\mathcal{O}(\Delta)$ values of $\gamma(H,k)$ at each of the $m-1$ internal vertices of the binary tree decomposition of $G$, we find we need $\mathcal{O}(m\Delta^2 \log \Delta)$ bits to store all values necessary to compute $|ET(G)|$.

The bounds above on the time and space complexity of computing $|ET(G)|$ are those given in Theorem 1.

## 5.3 Sampling Euler tours

We now show how to sample uniformly from $ET(G)$. We will sample a Euler tour by first sampling an $(s,t)$-decomposition, for some $k \in \gamma(G,k)$, and then applying Lemma 6 to obtain a uniform random tour. Note that after counting the number of Euler tours of $G$, we now know value of $\gamma_u(k) = \gamma(G_u, k)$ for every component graph in the tree decomposition.

$(s,t)$-decompositions are grouped in terms of the number of $(s,t)$-simple paths, $k$, in the decomposition. Therefore we first must choose $k \in \kappa(G)$, with the appropriate probability. Using Corollary 7, it is clear that the probability that an element of $ET(G)$ is consistent with an $(s,t)$-decomposition with $k$ $(s,t)$-simple paths, for any given $k \in \kappa(G)$, is exactly

$$\frac{\binom{k}{\frac{k}{2}}\frac{k}{4}2^{-k}\gamma(G,k)}{\sum_{l\in\kappa(G)}\binom{l}{\frac{l}{2}}\frac{l}{4}2^{-l}\gamma(G,l)}.$$

Our sampling algorithm computes these probabilities for all $k \in \kappa(G)$, and then chooses $k \in \kappa(G)$ to be an exact sample from this known distribution.

Having fixed $k$ for $G$, we must choose $k_1$ and $k_2$ values for the graphs $G_1$ and $G_2$ that combine (either via $o_s$ or $o_p$ or $o_d$) to form $G$. Suppose $G = G_1 o_s G_2$. We require $(k_1, k_2)$ such that $k_1, k_2 \geq k$. The probability with which we pick any $(k_1, k_2)$ such that $k_i \in \kappa(G_i)$ and $k_i \geq k$ for $i = 1, 2$ is exactly

$$\frac{\gamma(G_1,k_1)*\gamma(G_2,k_2)*\frac{k_1!k_2!}{k!\frac{k_1-k}{2}!\frac{k_2-k}{2}!}*\frac{\left(\frac{d(t_1)+d(s_2)}{2}-1\right)!}{\left(\frac{k_1+k_2}{2}-1\right)!}*\frac{2^{(d(t_1)+d(s_2))/2}}{2^{k_1+k_2-k}}}{\gamma(G,k)}. \tag{3}$$

For the case of $o_p$, the probability of choosing any $(k_1, k-k_1)$ for $k_1 \in \kappa(G_1), k-k_1 \in \kappa(G_2)$ is

$$\frac{\gamma(G_1,k_1)*\gamma(G_2,k-k_1)}{\gamma(G,k)}. \tag{4}$$

For the case of $o_d$, the probability of choosing any $(k, k_2)$ for $k \in \kappa(G_1), k_2 \in \kappa(G_2)$ is

$$\frac{\gamma(G_1,k)*\gamma(G_2,k_2)*\frac{k_2!}{\frac{k_2}{2}!(\frac{k_2}{2}-1)!}*\left(\frac{d_{G_2}(t')}{2}-1\right)!*2^{d_{G_2}(t')/2-k_2}}{\gamma(G,k)}. \tag{5}$$

Having chosen $k_i$, we recurse in the respective sub-trees and use (3), (4) or (5) to choose $k$ values with the appropriate probability for every internal node in the binary decomposition tree. Then for every $u \in T(G)$, we have generated a value $k_u$ from the exact uniform distribution for the $(s_u, t_u)$-decomposition of $G_u$. Next we must generate the $(s_u, t_u)$-decompositions themselves.

Suppose at node $u$ we have $G_u = G_1 o_p G_2$ and we have $C_i$ the respective $(s_i, t_i)$-decompositions for graph $G_i$, $i \in \{1,2\}$. Note that $s_1 \sim s_2$ and $t_1 \sim t_2$. We need to construct the $(s_u, t_u)$-decomposition for $G_u$. From Lemma 8, it is clear that $(C_1, C_2)$ is an $(s_u, t_u)$-decomposition for $G_u$.

Suppose at node $u$ we have $G_u = G_1 o_s G_2$. Let $C_i$ be the $(s_i, t_i)$-decompositions generated for graph $G_i$, $i \in \{1,2\}$, where $C_i$ is a exact uniform sample from $\mathcal{C}(G_i, k_i)$ for the fixed $k_i$ values, $i = 1, 2$. Note that $t_1 \sim s_2, s_u \sim s_1$ and $t_u \sim t_2$. Then by Lemma 10, we can combine $C_1$ and $C_2$ in a suitable fashion to obtain an $(s_u, t_u)$-decomposition for $G_u$. We have to ensure that every $(s_u, t_u)$- decomposition with $k$ $(s_u, t_u)$ simple paths which have both $s_u$ and $t_u$ as end-points that could arise from the combination of the given decompositions of $G_1$ and $G_2$ is equally likely. By Lemma 10, we know that we can construct a random $(s_u, t_u)$-decomposition of $G_u$ from $C_1, C_2$ by generating a random tuple of the following form:

$$\left(\pi_1\{1,\ldots,k_1\}, \pi_2\{1,\ldots,k_2\}, x \in \mathbb{N}_0^{(k_1+k_2)/2}, \sigma\{1,\ldots,\frac{D(k_1,k_2)}{2}\}, \{1,-1\}^{D(k_1,k_2)/2}\right),$$

such that the vector $x$ sums to $D(k_1, k_2)/2$, and then following the steps described in the proof of Lemma 10. This random tuple can easily be generated in polynomial-time. Hence we have the $o_s$ case.

The $(s_u, t_u)$-decomposition in the dangling case can also be constructed by first recursing on the (smaller) generalized series-parallel graph in a similar fashion as above. (Recall that we know that this smaller graph is itself Eulerian.) We use this decomposition of the subgraph to form a set of $(s_u, s_u)$ loops that we add to the decomposition for the parent node (in the tree) using the method described in Lemma 12.

The last and final step is to combine the $(s, t)$-decomposition at the root node of the binary decomposition tree into an Euler tour where $s$ and $t$ are the terminals of the generalized series-parallel graph. This can again be done simply, and in polynomial-time, by generating a random tuple of the form described in Lemma 6 and following the steps in the proof of that Lemma.

### 5.4 Other Counting Problems in GSP graphs

We wish to point out that the following list of combinatorial structures could be counted exactly using the above technique:(i) independent sets, (ii) matchings, (iii) $k$-colourings ($k$-constant), (iv) dominating sets. At the same time, we do not lay claim to being the first to do so. In fact, a recent result by Steve Noble [9] shows that (i),(ii) and (iii) could be counted exactly for a larger class of graphs, namely, the class of bounded treewidth graphs.